\documentclass[%
 reprint,
 amsmath,amssymb,a
 aps,
]{revtex4-2} 

\usepackage{graphicx}
\usepackage{dcolumn}
\usepackage{bm}
\usepackage[colorlinks,linkcolor=blue,citecolor=blue]{hyperref}
\usepackage{graphicx}

\usepackage{url}
\usepackage{ulem}
\usepackage{amsmath}
\usepackage{amsfonts}
\usepackage{textcomp}
\usepackage{subfigure}
\usepackage{verbatim}
\usepackage{rotating}

\usepackage{multirow} 
\usepackage{booktabs}
\usepackage{lineno}

\def \mev  {~\mbox{MeV}}

\def \jpsi {J/\psi}
\def \G {\gamma}
\def \GG {\gamma\gamma}
\def \GGG {\gamma\gamma\gamma}
\def \pipi {\pi^+\pi^-}
\def \kev {~\rm{keV}}

\begin{document}

\title{Observation of the charmonium decay $\eta_c\to\GG$ in $J/\psi\to\gamma\eta_c$}

\author{
M.~Ablikim$^{1}$, M.~N.~Achasov$^{4,c}$, P.~Adlarson$^{76}$, X.~C.~Ai$^{81}$, R.~Aliberti$^{35}$, A.~Amoroso$^{75A,75C}$, Q.~An$^{72,58,a}$, Y.~Bai$^{57}$, O.~Bakina$^{36}$, Y.~Ban$^{46,h}$
% ...
\\
\vspace{0.2cm}
(BESIII Collaboration)\\
\vspace{0.2cm} {\it
$^{1}$ Institute of High Energy Physics, Beijing 100049, People's Republic of China\\
% ...
$^{4}$ Budker Institute of Nuclear Physics SB RAS (BINP), Novosibirsk 630090, Russia\\
% ...
$^{35}$ Johannes Gutenberg University of Mainz, Johann-Joachim-Becher-Weg 45, D-55099 Mainz, Germany\\
$^{36}$ Joint Institute for Nuclear Research, 141980 Dubna, Moscow region, Russia\\
% ...
$^{46}$ Peking University, Beijing 100871, People's Republic of China\\
% ...
$^{57}$ Southeast University, Nanjing 211100, People's Republic of China\\
$^{58}$ State Key Laboratory of Particle Detection and Electronics, Beijing 100049, Hefei 230026, People's Republic of China\\
% ...
$^{72}$ University of Science and Technology of China, Hefei 230026, People's Republic of China\\
% ...
$^{75}$ University of Turin and INFN, (A)University of Turin, I-10125, Turin, Italy; (B)University of Eastern Piedmont, I-15121, Alessandria, Italy; (C)INFN, I-10125, Turin, Italy\\
$^{76}$ Uppsala University, Box 516, SE-75120 Uppsala, Sweden\\
% ...
$^{81}$ Zhengzhou University, Zhengzhou 450001, People's Republic of China\\

\vspace{0.2cm}
$^{a}$ Deceased\\
% ...
$^{c}$ Also at the Novosibirsk State University, Novosibirsk, 630090, Russia\\
% ...
$^{h}$ Also at State Key Laboratory of Nuclear Physics and Technology, Peking University, Beijing 100871, People's Republic of China\\
% ...
}

}


\begin{abstract}
Using $(2712.4\pm14.3)\times10^{6}$ $\psi(3686)$ events collected with the BESIII detector at the BEPCII collider, the decay $\eta_c\to\GG$ in $J/\psi\to\gamma\eta_c$ is observed. 
We determine the product branching fraction $\mathcal{B}(\jpsi\to\G\eta_c)\times\mathcal{B}(\eta_c\to\GG)=(5.23\pm0.26_{\rm{stat.}}\pm0.30_{\rm{syst.}})\times10^{-6}$. This result is consistent with the LQCD calculation $(5.34\pm0.16)\times10^{-6}$ from HPQCD in 2023.
By using the world-average values of $\mathcal{B}(\jpsi\to\G\eta_c)$ and the total decay width of $\eta_c$, the partial decay width $\Gamma(\eta_c\to\GG)$ is determined to be $(11.30\pm0.56_{\rm{stat.}}\pm0.66_{\rm{syst.}}\pm1.14_{\rm{ref.}})\kev$, which deviates from the corresponding world-average value by $3.4\sigma$.
\end{abstract}

  

\oddsidemargin  -0.2cm
\evensidemargin -0.2cm
\maketitle

\begin{figure}[htbp] \centering
	\setlength{\abovecaptionskip}{-1pt}
	\setlength{\belowcaptionskip}{10pt}

        \subfigure[]
        {\includegraphics[width=0.23\textwidth]{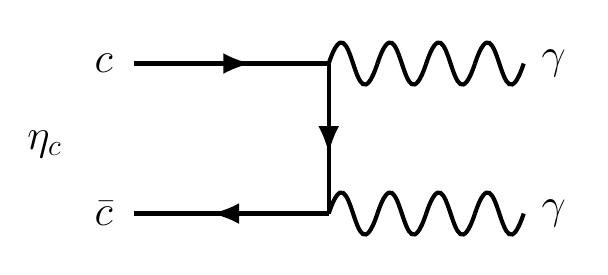}}
        \subfigure[]
        {\includegraphics[width=0.23\textwidth]{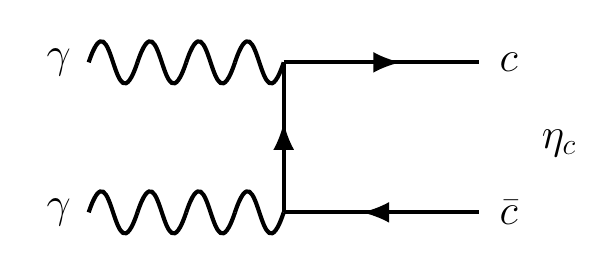}}\\
        
	\caption{The Feynman diagrams of the direct process $\eta_c\to\GG$ (a), and the time-inversion process $\gamma\gamma\to\eta_c$ (b).
    } 
	\label{fig:schematic_diagram}
\end{figure}

Charmonium systems offer a golden probe for investigating the nature of quantum chromodynamics (QCD), which is the fundamental theory governing strong interactions. Due to the medium energy scale of charmonium systems in strong interactions, charmonium physics encompasses both perturbative and non-perturbative phenomena~\cite{Brambilla:2010cs,Hagiwara:1980nv}, making it a valuable testing ground for deepening our understanding of QCD for both sides.
The decay rate of charmonium
can offer access to the strong coupling constant at the charmonium scale within the framework of perturbative QCD~\cite{Voloshin:2007dx}, and also provide a sensitive test to the application of the lattice QCD (LQCD) and the effective field theories such as non-relativistic QCD (NRQCD)~\cite{Bodwin:1994jh}.
Among these studies, $\eta_c\to\gamma\gamma$, as depicted in the Feynman diagram shown in Fig~\ref{fig:schematic_diagram}(a), has received significant theoretical attention.

The decay width of $\eta_c\to\GG$ can be written as $\Gamma(\eta_c\to\gamma\gamma)=\pi\alpha^2Q_c^4M_{\eta_c}F^2$~\cite{CLQCD:2020njc,CLQCD:2016ugl,Liu:2020qfz}, where $Q_c$ is the electric charge of the $c$ quark in units of $e$, $M_{\eta_c}$ is the mass of $\eta_c$, $\alpha$ is the fine structure constant, representing the electromagnetic interaction, and $F$ is the transition form factor, representing the strong interaction. To date, the electromagnetic part of the decay amplitude has been well understood, while the strong interaction part needs further study.
Within the framework of NRQCD, the relationship between the partial widths of $\jpsi\to e^+e^-$ and $\eta_c\to\GG$ can be expressed as $\mathcal R=\frac{\Gamma(\jpsi\to e^+e^-)}{\Gamma(\eta_c\to\GG)}=\frac{1}{3Q_c^2}(1+\mathcal{O}(\alpha_s)+\mathcal{O}(v^2/c^2))\approx\frac{3}{4}$ at the leading order (LO)~\cite{Czarnecki:2001zc}, where $\alpha_s$ is the running strong coupling constant, $v$ is the quark velocity in the charmonium system and $c$ is the velocity of light. It is expected to receive sizeable radiative and relativistic corrections by QCD~\cite{Czarnecki:2001zc,Bodwin:2001pt,Brambilla:2018tyu}, and their contribution may cause a deviation from $\frac{3}{4}$ for $\mathcal R$.
However, the phenomenology of $\eta_c\to\gamma\gamma$ for both experiment~\cite{CLEO:2008qfy,BESIII:2012lxx,pdg:2024} and theory~\cite{Czarnecki:2001zc,Bodwin:2001pt,Brambilla:2018tyu,Yu:2019mce,CLQCD:2020njc,Dudek:2006ut,CLQCD:2016ugl,Chen:2016bpj,Li:2019ncs,Liu:2020qfz,Zhang:2021xvl,Feng:2017hlu,Meng:2021ecs,Colquhoun:2023zbc} are still unclear until now.

Experimentally, the partial decay of $\eta_c\to \gamma\gamma$ could be assessed via radiative charmonium decays, $p\bar{p}$ annihilation (called direct processes), and the two-photon fusion process followed by hadrons (called time-inversion process), with Feynman diagrams shown in Fig~\ref{fig:schematic_diagram}(a) and Fig~\ref{fig:schematic_diagram}(b), respectively.
Evidence for $\eta_c\to\GG$ was obtained in $J/\psi\to\gamma\eta_c$ at both CLEO and BESIII, giving the product branching fraction (BF) $(1.2^{+2.7}_{-1.1}\pm0.3)\times 10^{-6}$~\cite{CLEO:2008qfy} and $(4.5\pm1.2\pm0.6)\times 10^{-6}$~\cite{BESIII:2012lxx} of $J/\psi\to\gamma\eta_c$, $\eta_c\to\GG$, respectively. 
Both results are consistent with each other but with large uncertainties. 
For another direct measurement, $p\bar{p}$ annihilation experiments have measured the cross-section for $p\bar{p}\to\gamma\gamma$ at various energy points and observed a peak around $\eta_c$ resonance~\cite{AnnecyLAPP:1987qkd,E760:1995rep,FermilabE835:2003ula}. This provides the product BF of $\mathcal{B}(\eta_c\to p\bar{p})\times\mathcal{B}(\eta_c\to\gamma\gamma)=(2.6\pm0.5)\times10^{-5}$~\cite{pdg:2024}.
Compared to the direct process, the measurements with the time-inverse process are more precise currently~\cite{Belle:2018bry,Belle:2007qae,Belle:2012qqr,BaBar:2011gos,BaBar:2010siw,TASSO:1988utg}, and a global fit gives a world-average value of ~$\Gamma(\eta_c\to\GG)=(5.1\pm0.4)\kev$~\cite{pdg:2024}. However, there are large discrepancies between different measurements of $\gamma\gamma\to\eta_c$~\cite{Colquhoun:2023zbc}.

\vspace{-0.0cm}
\vspace{-0.0cm}
There are a number of theoretical predictions for $\Gamma(\eta_c\to\GG)$~\cite{Czarnecki:2001zc,Bodwin:2001pt,Brambilla:2018tyu,Yu:2019mce,CLQCD:2020njc,Dudek:2006ut,CLQCD:2016ugl,Chen:2016bpj,Li:2019ncs,Liu:2020qfz,Zhang:2021xvl,Feng:2017hlu,Meng:2021ecs,Colquhoun:2023zbc}, which do not agree with each other and with the world-average value.
The LQCD calculation, which employs a model-independent approach, deviates from the world-average value by more than $3\sigma$~\cite{Meng:2021ecs}, while the LQCD calculation by HPQCD, which for the first time includes $u$, $d$, $s$, and $c$ quarks in the sea, shows a tension exceeding $4\sigma$~\cite{Colquhoun:2023zbc}. Additionally, a next-to-next-to-leading order (NNLO) QCD correction for $\Gamma(\eta_c\to\GG)$ suggests that the calculated result is substantially larger than the world-average value by more than $10\sigma$~\cite{Feng:2017hlu}. 
These deviations are all in the same direction and 
there may be aspects that require further understanding.
Therefore, a new and precise measurement of $\eta_c\to\GG$, which is independent of the $p\bar{p}$ annihilation and the two-photon production experiments, is crucial for resolving this issue.

In this Letter, we present the measurement of $\eta_c\to\GG$ via $\psi(3686)\to \pi^+\pi^-J/\psi$ with $J/\psi\to \gamma\eta_c$ using $(2712.4\pm14.3)\times10^{6}$ $\psi(3686)$ events~\cite{BESIII:2017tvm,BESIII:2024lks} collected with the BESIII detector at the BEPCII collider. This sample has a particular advantage over using a directly produced $J/\psi$ sample due to the high $e^+e^-\to\gamma_{\rm{ISR}}\gamma\gamma$ background in the latter sample.

The BESIII detector~\cite{BESIII:2009fln} records symmetric $e^+e^-$ collisions 
provided by the BEPCII storage ring~\cite{Yu:2016cof}
in the center-of-mass energy ($\sqrt s$) range from 1.85 to 4.95~GeV,
with a peak luminosity of $1.1 \times 10^{33}\;\text{cm}^{-2}\text{s}^{-1}$ 
achieved at $\sqrt{s} = 3.773\;\text{GeV}$. 
The cylindrical core of the $\mbox{BESIII}$ detector covers 93\% of the full solid angle and consists of a helium-based multilayer drift chamber~(MDC), a plastic scintillator time-of-flight system~(TOF), and a CsI(Tl) electromagnetic calorimeter~(EMC), which are all enclosed in a superconducting solenoidal magnet providing a 1.0~T magnetic field (0.9~T in year 2012).
The solenoid is supported by an octagonal flux-return yoke with resistive plate counter muon identification modules interleaved with steel. 
The charged-particle momentum resolution at $1~{\rm GeV}/c$ is $0.5\%$, and the ${\rm d}E/{\rm d}x$
resolution is $6\%$ for electrons from Bhabha scattering. The EMC measures photon energies with a resolution of $2.5\%$ ($5\%$) at $1$~GeV in the barrel (end cap) region. The time resolution in the TOF barrel region is 68~ps, while that in the end cap region was 110~ps. The end cap TOF system was upgraded in 2015 using multigap resistive plate chamber technology, providing a time resolution of
60~ps, which benefits 83.3\% of the data used in this analysis~\cite{Cao:2020ibk,Li:2017jpg,Guo:2017sjt}.

Simulated data samples, generated with a {\sc
geant4}-based~\cite{GEANT4:2002zbu} Monte Carlo (MC) package~\cite{ABLIKIM2010345} that includes the geometric description of the BESIII detector~\cite{Huang:2022wuo,Liang:2009zzb,Zheng-Yun_2008} and the detector response, are utilized to determine detection efficiencies and to estimate backgrounds. The simulation models the beam energy spread and initial state radiation in $e^+e^-$ annihilations with the generator {\sc
kkmc}~\cite{Jadach:2000ir,Jadach:1999vf}.
All particle decays are modeled with {\sc
evtgen}~\cite{Lange:2001uf,Ping:2008zz}, using BFs either taken from the Particle Data Group (PDG)~\cite{pdg:2024} when available, or otherwise estimated with {\sc lundcharm}~\cite{Chen:2000tv,Yang:2014vra}.
Final state radiation from charged final state particles is incorporated using the {\sc photos} package~\cite{Barberio:1990ms}.

The selection criteria of $\psi(3686)\to\pipi\jpsi$, $\mbox{$\jpsi\to\gamma\eta_c$}$, $\eta_c\to\GG$ are described below. 
Two charged pion tracks are required to be within a polar angle ($\theta$) range of $|\rm{\cos\theta}|<0.93$, where $\theta$ is defined with respect to the $z$ axis, which is the symmetry axis of the MDC. For both pions, the distance of the closest approach to the interaction point (IP) must be less than 10\,cm along the $z$ axis and less than 1\,cm in the transverse plane.
Photon candidates are identified using showers in the EMC.  The deposited energy of each shower must exceed 50~MeV in the barrel region ($|\cos \theta|< 0.80$) or in the end cap region ($0.86 <|\cos \theta|< 0.92$). 
To suppress electronic noise and showers unrelated to the event, the difference between the EMC shower time and the event start time~\cite{Guan:2013jua} is required to be within [0, 700]\,ns.
In addition, to exclude showers that originate from charged tracks, the angle subtended by the EMC shower and the position of the closest charged track at the EMC must be greater than 10 degrees as measured from the IP. 
Only the events with two charged pion tracks and three or four photons are selected~\cite{Li:2024pox}. 
To suppress the background from \mbox{$e^+e^-\to e^+e^-\GGG$}, the combined likelihoods ($\mathcal{L}$) of particle identification (PID) under the positron and pion hypotheses are obtained, and pion candidates are required to satisfy $\mathcal{L}(e)/(\mathcal{L}(e)+\mathcal{L}(\pi))<0.8$.
A four-constraint kinematic fit~\cite{Yan:2010zze}, constraining the total four-momentum of the final state to the initial state is performed. If multiple combinations of $\pipi\GGG$ are found, the one with the minimal $\chi^2_{\rm{4C}}$ value is selected. We require $\chi^2_{\rm{4C}}<19$, determined by optimizing the signal sensitivity $\frac{S}{\sqrt{S+B}}$, where $S$ is the expected signal yield and $B$ is the expected background yield.
To select the $\jpsi$ candidate, the recoil mass of $\pipi$, defined as $\mbox{$M^{\rm{recoil}}_{\pipi}=|\textbf{p}_{e^+} + 
\textbf{p}_{e^-} - \textbf{p}_{\pi^+} - \textbf{p}_{\pi^-}|$}$ with the four-momenta of the beam particles $\textbf{p}_{e^+}$ and $\textbf{p}_{e^-}$ and the reconstructed four-momentum $
\textbf{p}_{\pi^+(\pi^-)}$ of $\pi^+(\pi^-)$, is required to be in the range $[3.083,~3.111]~\rm{GeV}/\it{c}^{\rm{2}}$.
The three selected photons are marked as $\gamma_1,~\gamma_2,~\gamma_3$ sorted by energy from highest to lowest. 
After applying all selection criteria, the dominant remaining backgrounds arise from the processes \mbox{$\jpsi\to\gamma\pi^0/\eta/\eta'\to\GGG$}.
While the $\eta_c$ signal is expected to be in the $\gamma_1\gamma_2$ invariant mass $M_{12}$, these backgrounds peak in the other two combinations and can be suppressed by rejecting the events with invariant mass of $\gamma_2\gamma_3$ ($M_{23}$) or $\gamma_1\gamma_3$ ($M_{13}$) falling in the regions \mbox{[0.10, 0.16]}, [0.48, 0.59], and [0.88, 1.10] GeV/$\it{c}^{\rm{2}}$.
An inverted veto region of $M_{23}$ and $M_{13}$ has been implemented to verify the background simulation, which shows good agreement with the data.
In addition, there are backgrounds from \mbox{$\jpsi\to\G\pi^0\pi^0\to\GG\GGG$} and $\jpsi\to\GGG$, as well as other minor contributions modeled with the $\psi(3686)$  inclusive MC sample~\cite{Zhou:2020ksj}.
The non-$\jpsi$ (non-$\psi(3686)$) background is estimated using the sideband region of $M^{\rm{recoil}}_{\pi^+\pi^-}$ (using the data samples at 3.773 GeV~\cite{BESIII:2024lbn}) and is found to be negligible.
The signal efficiency is estimated to be $\epsilon_{\rm{sig}}=(13.77\pm0.02)\%$ from signal MC simulation.

        

\vspace{-0.0cm}
\begin{figure}[htbp] \centering
	\setlength{\abovecaptionskip}{-1pt}
	\setlength{\belowcaptionskip}{10pt}

        {\includegraphics[width=0.49\textwidth]{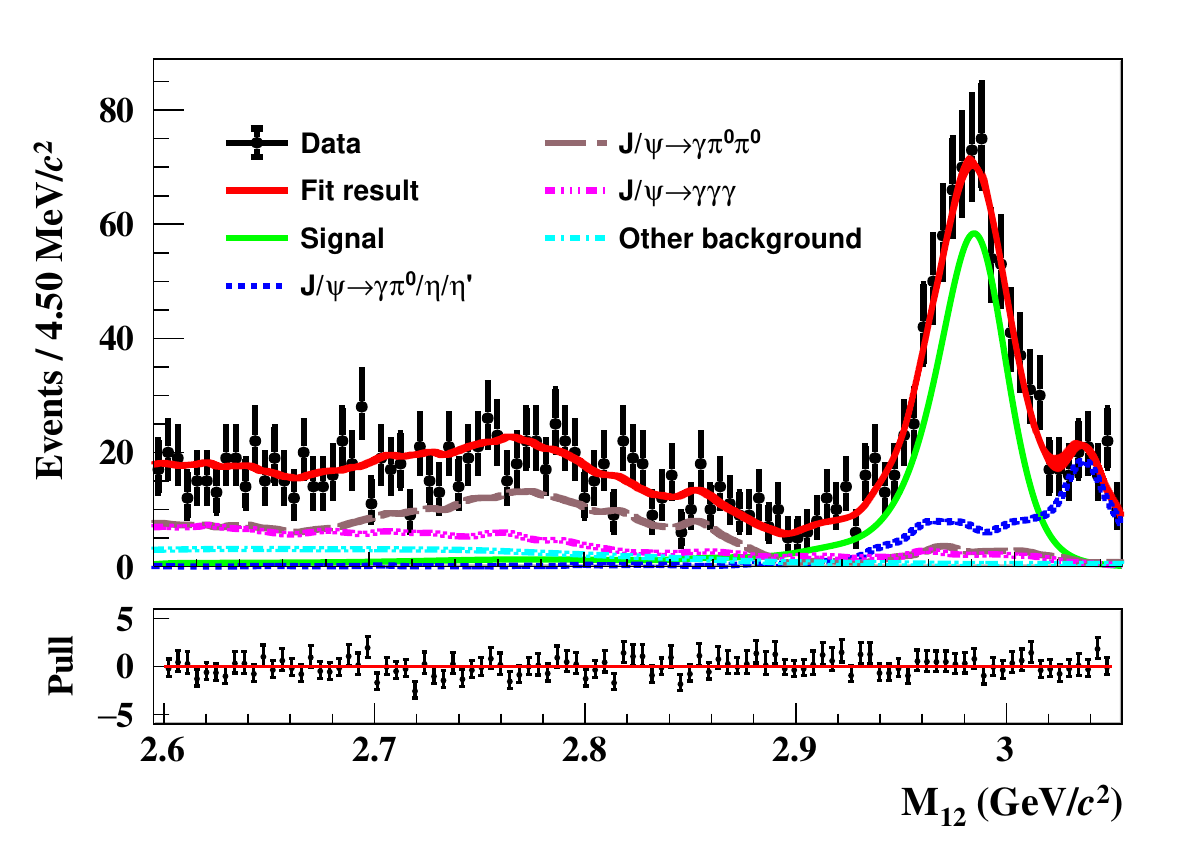}}
        
	\caption{Fit to the $M_{12}$ distribution. The black points with error bars are data, the red line is the fit result, the green line is the signal, and the lines in other colors represent different backgrounds.
    } 
	\label{fig:fit}
\end{figure}
\vspace{-0.0cm}

Following the application of the aforementioned selection criteria, a significant $\eta_c\to\GG$ peak in the $M_{12}$ distribution is observed, as shown in Fig~\ref{fig:fit}. 
To extract the signal yield of $\eta_c\to\GG$, an unbinned extended maximum likelihood fit on this distribution is performed.
In the fit, the lineshape of $\eta_c$ is described as
\begin{eqnarray}
\mathcal{F}(m)=E_{\gamma}^3(m)\times f_{\rm{damp}}(m)\times |BW(m)|^2
\end{eqnarray}
with
\begin{eqnarray}
BW(m)=\frac{M_{\eta_c}\Gamma_{\eta_c}}{m^2-M_{\eta_c}^2+iM_{\eta_c}\Gamma_{\eta_c}},
\end{eqnarray}
\begin{eqnarray}
E_{\gamma}(m)=\frac{M_{\jpsi}^2-m^2}{2\times M_{\jpsi}},
\end{eqnarray}
and
\begin{eqnarray}
f_{\rm{damp}}(m)=\frac{E_{0}^2}{E_0 \times E_{\gamma}(m)+(E_0-E_{\gamma}(m))^2},
\end{eqnarray}
where $m$ is the mass of $\eta_c$, $E_{\gamma}^3(m)$ is the M1 transition form factor~\cite{Brambilla:2005zw}, $E_{\gamma}(m)$ is the energy of the transition photon in the rest frame of $\jpsi$, $f_{\rm{damp}}(m)$ is the damping factor in reducing the long tail from the M1 transition form factor~\cite{CLEO:2008pln,Anashin:2010nr}, $M_{\eta_c}$($\Gamma_{\eta_c}$) is the nominal mass (width) of $\eta_c$~\cite{pdg:2024}, which is assumed to follow a Gaussian distribution and constrained with the PDG values~\cite{pdg:2024} (Gaussian constraint), $M_{\jpsi}$ is the nominal mass of $\jpsi$~\cite{pdg:2024} and $E_{0}=\frac{M_{\jpsi}^2-M_{\eta_c}^2}{2\times M_{\jpsi}}$ is the energy of the transition photon at the peak of $\eta_c$.
The signal probability density function in the fit is described by
\begin{eqnarray}
\mathcal{PDF}(m)\sim[\epsilon(m)\times\mathcal{F}(m)]\otimes G(\mu,\sigma),
\end{eqnarray}
where $\epsilon(m)$ is the mass-dependent efficiency studied from the signal MC simulation, $G(\mu,\sigma)$ is a Gaussian function with free parameters $\mu$ and $\sigma$ to account for the detector resolution. The background shapes are extracted from the $\psi(3686)$ inclusive MC sample with kernel density estimation~\cite{Poluektov:2014rxa}. The strengths of each background component also need to satisfy the Gaussian constraint given by the luminosity and the known BFs of these background processes~\cite{pdg:2024}. The resultant fit result is shown in Fig~\ref{fig:fit} and the fit gives the signal yield $N_{\rm{sig}}=677.7\pm33.5$.

The product BF of $\jpsi\to\G\eta_c$ and $\eta_c\to\GG$ is calculated by 
\begin{equation}
\begin{aligned}
	\label{eq:BF}
	&\mathcal{B}(\jpsi\to\G\eta_c)\times\mathcal{B}(\eta_c\to\GG)\\
	&=\frac{N_{\rm{sig}}}{N_{\psi(3686)}\times\epsilon_{\rm{sig}}\times\mathcal{B}(\psi(3686)\to\pipi\jpsi)},
\end{aligned}
\end{equation}
where $N_{\psi(3686)}$ is the total number of $\psi(3686)$ events~\cite{BESIII:2017tvm,BESIII:2024lks}, $\epsilon_{\rm{sig}}$ is the signal efficiency studied from the signal MC simulation, and $\mathcal{B}(\psi(3686)\to\pipi\jpsi)$ is adopted from the PDG~\cite{pdg:2024}.
The number of $\eta_c\to\gamma\gamma$ events from non-``$\pi^+\pi^-J/\psi$, $J/\psi\to\gamma\eta_c$" decay of $\psi(3686)$ is negligible due to their relatively small BF and low detection efficiency.
The decay width of $\eta_c\to\GG$ is calculated by 
\begin{equation}
\begin{aligned}
	\label{eq:Gamma_GG}
	\Gamma(\eta_c\to\GG)
	&=\frac{\mathcal{B}(\jpsi\to\G\eta_c)\times\mathcal{B}(\eta_c\to\GG)}{\mathcal{B}^{\rm{PDG}}(\jpsi\to\G\eta_c)}\times\Gamma^{\rm{PDG}}_{\eta_c},
\end{aligned}
\end{equation}
where ${\mathcal B}^{\rm PDG}(J/\psi \to \gamma \eta_c)$ is the BF of $\jpsi\to\G\eta_c$ and $\Gamma^{\rm{PDG}}_{\eta_c}$ is the total width of $\eta_c$, with both values obtained from the PDG~\cite{pdg:2024}.

The sources of systematic uncertainty include the total number of $\psi(3686)$ events, intermediate BF, signal efficiency, and signal extraction. The uncertainty of the total number of $\psi(3686)$ events is 0.5\%~\cite{BESIII:2024lks}. The uncertainty of the BF of $\psi(3686)\to\pipi\jpsi$ is 1.0\%~\cite{pdg:2024}. The uncertainty of the tracking for two pions is assigned as 0.6\% from the control sample $\jpsi\to\pipi\pi^0$.
The uncertainty of photon detection is found to be 0.5\% per photon from the control samples $e^+e^-\to\gamma\mu^+\mu^-$.
The uncertainty associated with  the $M_{23}$ and $M_{13}$ veto selection is studied by smearing the MC-generated $M_{23}$ and $M_{13}$ spectra with a Gaussian function $G'(\mu',\sigma')$, setting $\mu'=\pm5\mev$ and $\sigma'=10\mev$. The maximum signal efficiency difference, 0.7\%, is taken as a conservative estimation of the uncertainty.
The uncertainties of other selections are estimated from the control sample $\psi(3686)\to\pipi\jpsi$, $\jpsi\to\gamma\eta$, $\eta\to\GG$. For each case, the uncertainty is taken as the efficiency difference between data and MC samples. The assigned uncertainty is 0.5\% for the PID of the two pions, 0.2\% for the $M^{\rm{recoil}}_{\pipi}$ requirement, 0.8\% for the photon number ($N_{\gamma}$) requirement, and 2.9\% for the $\chi^2_{\rm{4C}}$ requirement.
The uncertainty of the $\eta_c$ lineshape is estimated by varying the damping factor from $f_{\rm{damp}}(m)=\frac{E_{0}^2}{E_0 \times E_{\gamma}(m)+(E_0-E_{\gamma}(m))^2}$~\cite{Anashin:2010nr} to \mbox{$f_{\rm{damp}}(m)=\rm{exp}(\frac{\it{E}_{\gamma}^{\rm{2}}(\it{m})}{8\beta^2})$}
with $\beta=65\mev$~\cite{CLEO:2008pln}, and the BF difference, 3.6\%, is assigned as the uncertainty.
The uncertainty of the background shape is investigated by changing the background shape with different kernel width parameters in kernel density estimation~\cite{Poluektov:2014rxa}, and the BF difference, 0.2\%, is taken as the uncertainty.
The uncertainty of the background yield is assigned by individually removing the Gaussian constraints on the background component yields, and the maximum BF difference, 2.7\%, is taken as the uncertainty.
The total systematic uncertainty is calculated to be 5.8\% by summing up all sources in quadrature. All aforementioned systematic uncertainties are summarized in Table~\ref{tab:syst_err}. 
The reference uncertainty sources for the $\Gamma(\eta_c\to\GG)$ measurement are from ${\mathcal B}^{\rm PDG}(J/\psi \to \gamma \eta_c)$ and $\Gamma^{\rm PDG}_{\eta_c}$, which are 9.9\% and 1.6\%, respectively~\cite{pdg:2024}. The combined effect of these sources is 10.1\%.

\begin{table}[ht] 
	\centering
	\caption{Relative systematic uncertainties in the measurement of $\mathcal{B}(\jpsi\to\G\eta_c)\times\mathcal{B}(\eta_c\to\GG)$.}
	\begin{tabular}{*{2}{c}}
		\midrule \midrule
		Source &  Uncertainty (\%)\\
		\midrule
		Tracking & 0.6\\\specialrule{0em}{0.5pt}{0.5pt}
            PID & 0.5\\\specialrule{0em}{0.5pt}{0.5pt}
            Photon detection & 1.5\\\specialrule{0em}{0.5pt}{0.5pt}
		$M_{\pipi}^{\rm{recoil}}$ requirement & 0.2\\\specialrule{0em}{0.5pt}{0.5pt}
		$N_{\G}$ requirement & 0.8\\\specialrule{0em}{0.5pt}{0.5pt}
		$\chi_{\rm{4C}}^2$ requirement  & 2.9\\\specialrule{0em}{0.5pt}{0.5pt}
            Veto selection & 0.7\\\specialrule{0em}{0.5pt}{0.5pt}
            Lineshape of $\eta_c$ & 3.6 \\\specialrule{0em}{0.5pt}{0.5pt}
            Background shape & 0.2\\\specialrule{0em}{0.5pt}{0.5pt}
            Background yield & 2.7\\\specialrule{0em}{0.5pt}{0.5pt}
		Total number of $\psi(3686)$ events & 0.5\\\specialrule{0em}{0.5pt}{0.5pt}
		$\mathcal{B}(\psi(3686)\to\pipi\jpsi)$ & 1.0\\\specialrule{0em}{0.5pt}{0.5pt}
		\midrule
		Total & 5.8\\
		\midrule \midrule
	\end{tabular}
	\label{tab:syst_err}
\end{table}


\vspace{-0.0cm}
\begin{figure}[htbp] \centering
	\setlength{\abovecaptionskip}{-1pt}
	\setlength{\belowcaptionskip}{10pt}

        {\includegraphics[width=0.49\textwidth]{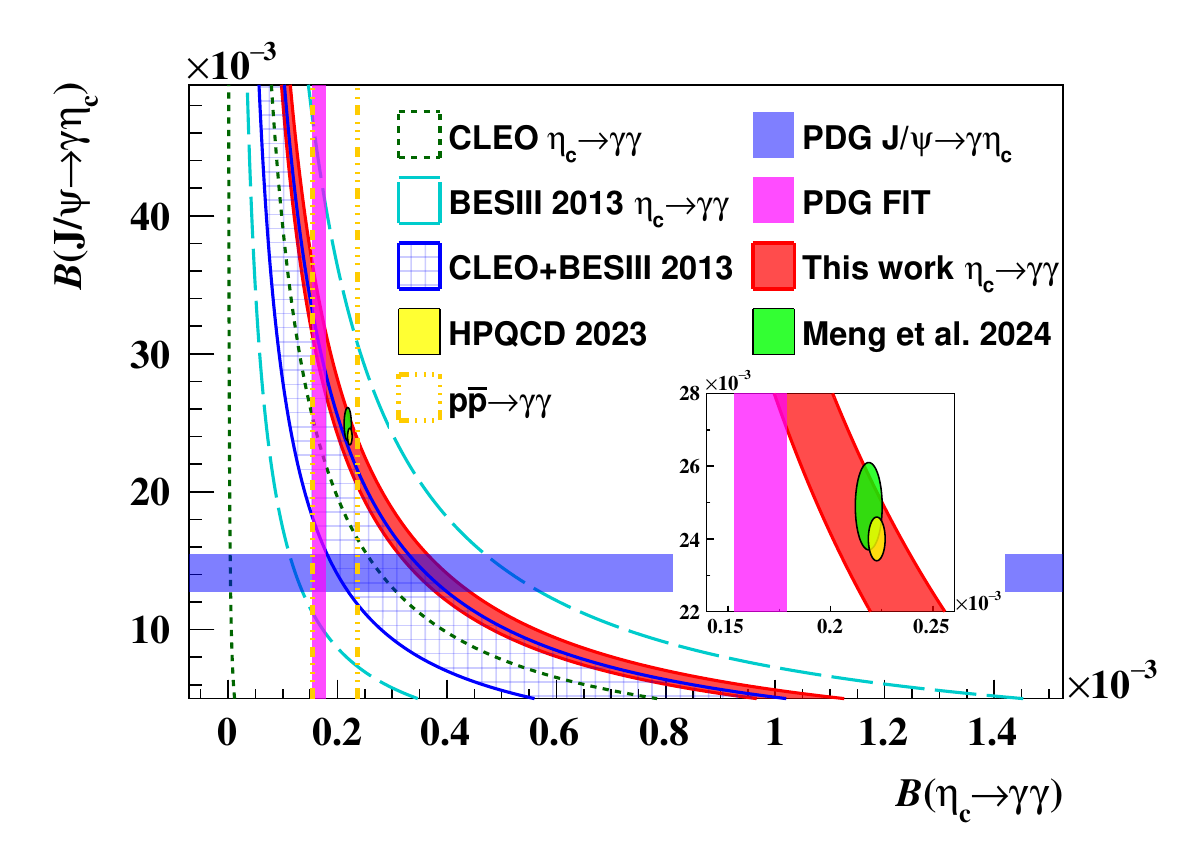}}
        
	\caption{The comparison of $\mathcal{B}(\eta_c\to\GG)$ versus \mbox{$\mathcal{B}(\jpsi\to\gamma\eta_c)$} with $1\sigma$ confidence level. The red-filled region is our result of $\mathcal{B}(\jpsi\to\G\eta_c)\times\mathcal{B}(\eta_c\to\GG)$, the yellow-filled region is the HPQCD calculation~\cite{Colquhoun:2023zbc} and the green-filled region is from the calculation of Meng et al.~\cite{Meng:2021ecs,Meng:2024axn}. The green line represents the CLEO result~\cite{CLEO:2008qfy}, the light blue line represents the previous \mbox{BESIII} result~\cite{BESIII:2012lxx}, and the blue grid-filled region is the combined result of CLEO and BESIII. The purple-red filled region is the $\mathcal{B}(\eta_c\to\GG)$ result from the PDG~\cite{pdg:2024} and the blue full-filled region is the $\mathcal{B}(\jpsi\to\gamma\eta_c)$ result from the PDG~\cite{pdg:2024}. The orange line represents the average $p\bar{p}\to\gamma\gamma$ result from SPEC~\cite{AnnecyLAPP:1987qkd}, E760~\cite{E760:1995rep}, and E835~\cite{FermilabE835:2003ula} normalized to $\mathcal{B}(\eta_c\to p\bar{p})=(1.33\pm0.11)\times10^{-3}$.
 }
	\label{fig:BF_sum}
\end{figure}
\vspace{-0.0cm}

The final product BF, \mbox{$\mathcal{B}(\jpsi\to\G\eta_c)\times\mathcal{B}(\eta_c\to\GG)$}, is calculated to be \mbox{$(5.23\pm0.26_{\rm{stat.}}\pm0.30_{\rm{syst.}})\times10^{-6}$} by Eq.~\ref{eq:BF}, where the first uncertainty is statistical, and the second is systematic.
A comparison of our result with previous measurements~\cite{CLEO:2008qfy,BESIII:2012lxx}, the world-average values~\cite{pdg:2024}, and LQCD calculations from HPQCD~\cite{Colquhoun:2023zbc} and Meng et al.~\cite{Meng:2021ecs,Meng:2024axn} in the $\mathcal{B}(\eta_c\to\GG)$ versus $\mathcal{B}(\jpsi\to\gamma\eta_c)$ plane is shown in Fig~\ref{fig:BF_sum}.
In the plot, the value from $p\bar{p}\to\gamma\gamma$ is normalized to $\mathcal{B}(\eta_c\to p\bar{p})=(1.33\pm0.11)\times10^{-3}$~\cite{pdg:2024}.
We find that the world-average values of $\mathcal{B}(\eta_c\to\GG)$ and $\mathcal{B}(\jpsi\to\gamma\eta_c)$~\cite{pdg:2024} do not simultaneously align with our measurement.
Interestingly, the highly precise LQCD predictions from HPQCD with $\mathcal{B}(\jpsi\to\G\eta_c)\times\mathcal{B}(\eta_c\to\GG)=(5.34\pm0.16)\times10^{-6}$ ~\cite{Colquhoun:2023zbc} 
 and Meng et al.~\cite{Meng:2021ecs,Meng:2024axn} both agree with our measurement, while the corresponding individual calculations of $\mathcal{B}(\eta_c\to\GG)$ and $\mathcal{B}(\jpsi\to\gamma\eta_c)$ are inconsistent with the world-average values~\cite{pdg:2024}. No other calculation provides both $\mathcal{B}(\eta_c\to\GG)$ and $\mathcal{B}(\jpsi\to\gamma\eta_c)$ simultaneously.

\vspace{-0.0cm}
\begin{figure}[htbp] \centering
	\setlength{\abovecaptionskip}{-1pt}
	\setlength{\belowcaptionskip}{10pt}

        {\includegraphics[width=0.49\textwidth]{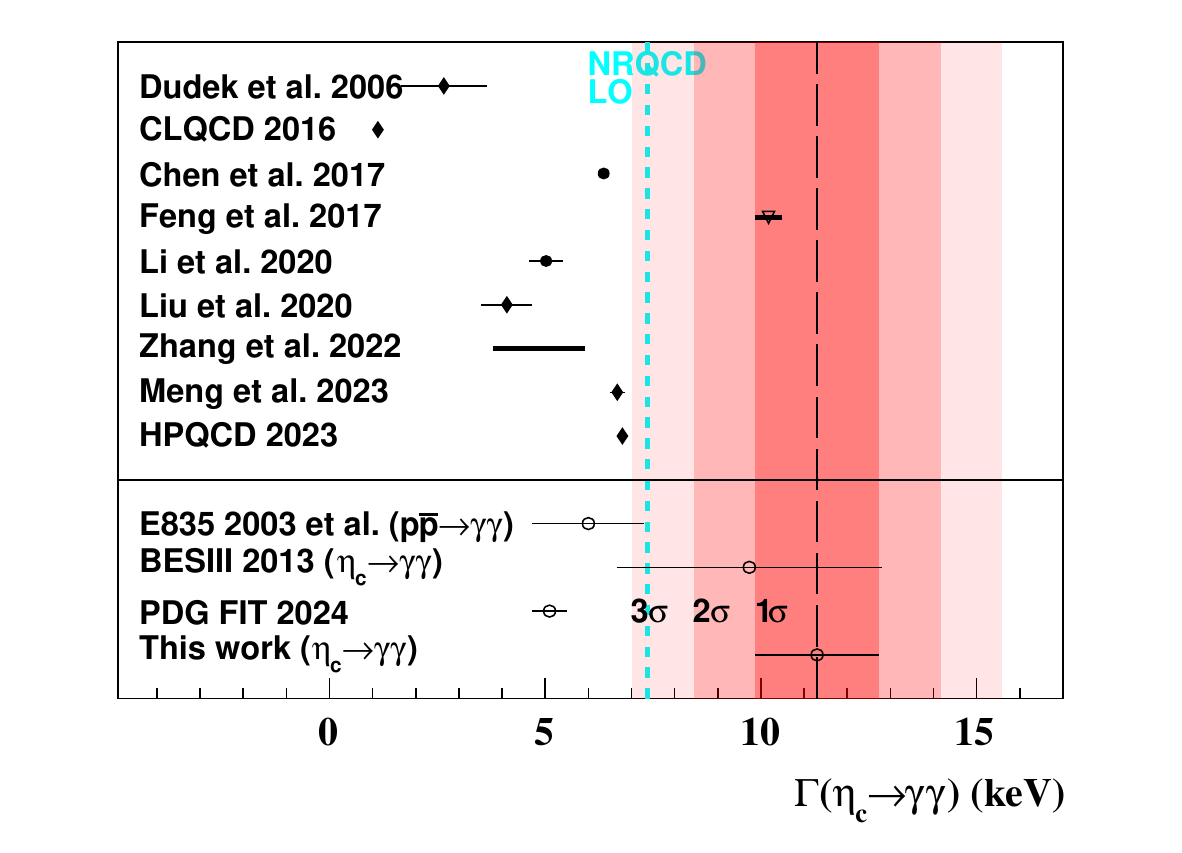}}
        
	\caption{The comparison of $\Gamma(\eta_c\to\GG)$. For the plot of ``This work", the black solid line indicates the total uncertainty including the reference uncertainties of $\jpsi\to\gamma\eta_c$ and $\Gamma_{\eta_c}$~\cite{pdg:2024}, and the dark, dark-to-light, and light red-filled regions indicate the $1\sigma$, $2\sigma$, and $3\sigma$ regions, respectively. The LQCD calculations are marked with rhomboid and the NNLO calculation of NRQCD is marked with an inverted triangle. The blue dashed line corresponds to $\mathcal{R}=\frac{3}{4}$ using $\Gamma(J/\psi\to e^+ e^-)$ from the PDG~\cite{pdg:2024}.
    } 
	\label{fig:Width_sum}
\end{figure}
\vspace{-0.0cm}

Using $\mathcal{B}(\jpsi\to\G\eta_c)=(1.41\pm0.14)\%$ and \mbox{$\Gamma_{\eta_c}=(30.5\pm0.5)\mev$} from the PDG~\cite{pdg:2024}, the decay width of $\eta_c\to\GG$ is determined to be \mbox{$\Gamma(\eta_c\to\GG)=(11.30\pm0.56_{\rm{stat.}}\pm0.66_{\rm{syst.}}\pm1.14_{\rm{ref.}})\kev$},
with the first uncertainties statistical, the second systematic, and the third from $\mathcal{B}(\jpsi\to\G\eta_c)$ and $\Gamma_{\eta_c}$ used from the PDG~\cite{pdg:2024}.
The decay widths $\Gamma(\eta_c\to\GG)$ from multiple theoretical calculations, experimental measurements, and the world-average values are shown in Fig~\ref{fig:Width_sum}.
Combining with the world-average value of $\mathcal{B}(\jpsi\to\G\eta_c)$~\cite{pdg:2024}, our result of ${\mathcal B}(\eta_c\to \gamma\gamma)$ is consistent with the NNLO corrections of NRQCD~\cite{Feng:2017hlu} but significantly deviates from other calculations by more than $3\sigma$.
Although the HPQCD calculation of $\mathcal{B}(\jpsi\to\G\eta_c)\times\mathcal{B}(\eta_c\to\GG)$~\cite{Colquhoun:2023zbc} is consistent with our measurement, it predicts a different value of $\mathcal{B}(\jpsi\to\G\eta_c)$ compared to the world-average value. This discrepancy is evident in the comparison between our measurement and the HPQCD value in Fig~\ref{fig:Width_sum}.
Our measurement also deviates from the world-average value of $\Gamma(\eta_c\to\GG)$~\cite{pdg:2024}, which is dominantly based on the measurements with the time-inverse process~\cite{Belle:2018bry,Belle:2007qae,Belle:2012qqr,BaBar:2011gos,BaBar:2010siw,TASSO:1988utg}, by $3.4\sigma$.
It is crucial to note that several theoretical calculations of $\mathcal{B}(\jpsi\to\gamma\eta_c)$ are significantly larger than the value presented in the PDG~\cite{Meng:2024axn,Colquhoun:2023zbc,Gui:2019dtm,Becirevic:2012dc,Donald:2012ga}.

In summary, we report the measurement of the charmonium decay $\eta_c\to\GG$ based on $(2712.4\pm14.3)\times10^{6}$ $\psi(3686)$ events. The reported product BF \mbox{$\mathcal{B}(\jpsi\to\G\eta_c)\times\mathcal{B}(\eta_c\to\GG)$}, \mbox{$(5.23\pm0.26_{\rm{stat.}}\pm0.30_{\rm{syst.}})\times10^{-6}$}, is consistent with the recent LQCD calculations~\cite{Colquhoun:2023zbc,Meng:2021ecs,Meng:2024axn}.
Using the world-average value of $\mathcal{B}(\jpsi\to\G\eta_c)$, we also present the decay width of $\eta_c\to\GG$ to be $(11.30\pm0.56_{\rm{stat.}}\pm0.66_{\rm{syst.}}\pm1.14_{\rm{ref.}})\kev$. It agrees with the NRQCD NNLO calculation~\cite{Feng:2017hlu} but deviates from the world-average value, which is dominantly based on the time-inverse process, by $3.4\sigma$.
When taking CPT conservation into account, the difference in the decay of $\eta_c\to\gamma\gamma$ compared to its time-inverse process may suggest a potential source of CP violation~\cite{Sakharov:1967dj}. However, before delving into this, a careful check of the current experimental results is necessary.
Our results indicate that the current experimental information of either $\eta_c\to\GG$, or $J/\psi\to\gamma\eta_c$ may not be fully reliable, underscoring the need for precise and independent measurements of both quantities in future studies. 
In both direct and time-inversion process measurements, the interference between the $\eta_c$-included process and the non-resonance process is neglected. Taking into account this potential interference with a full interference assumption~\cite{BESIII:2011ab}, the product BF $\mathcal{B}(\jpsi\to\G\eta_c)\times\mathcal{B}(\eta_c\to\GG)$ becomes $(4.13\pm0.20\pm0.23)\times10^{-6}$ for the constructive interference or $(6.73\pm0.32\pm0.38)\times10^{-6}$ for the destructive interference, with a statistical significance of only 1.2$\sigma$ by comparing the likelihood in the fits with and without interference.
Due to the limited data sample and the good quality of the fit without interference, we report the fit result without interference as the nominal result and provide the fit results with interference for reference.
Furthermore, given the relatively large BF uncertainty of $J/\psi \to \gamma \eta_c$, additional studies of $\eta_c \to \gamma\gamma$ produced through alternative mechanisms, such as $\psi(3686) \to \gamma \eta_c$ or $h_c \to \gamma \eta_c$, could offer further cross-validation of $\Gamma(\eta_c \to \gamma\gamma)$ with independent systematic uncertainties. Additionally, an updated measurement of $\mathcal{B}(J/\psi \to \gamma \eta_c)$ with $\eta_c\to\rm{inclusive}$ or $\eta_c\to\rm{hadrons}$ is also essential to further investigate the QCD phenomenon in charmonium.

\textbf{Acknowledgement}

The BESIII Collaboration thanks the staff of BEPCII and the IHEP computing center for their strong support. This work is supported in part by National Key R\&D Program of China under Contracts Nos. 2023YFA1606000, 2020YFA0406400, 2020YFA0406300; National Natural Science Foundation of China (NSFC) under Contracts Nos. 11635010, 11735014, 11935015, 11935016, 11935018, 12025502, 12035009, 12035013, 12061131003, 12175321, 12192260, 12192261, 12192262, 12192263, 12192264, 12192265, 12221005, 12225509, 12235017, 12361141819; the Chinese Academy of Sciences (CAS) Large-Scale Scientific Facility Program; the CAS Center for Excellence in Particle Physics (CCEPP); Joint Large-Scale Scientific Facility Funds of the NSFC and CAS under Contract No. U1832207, U1932101; 100 Talents Program of CAS; The Institute of Nuclear and Particle Physics (INPAC) and Shanghai Key Laboratory for Particle Physics and Cosmology; German Research Foundation DFG under Contracts Nos. FOR5327, GRK 2149; Istituto Nazionale di Fisica Nucleare, Italy; Knut and Alice Wallenberg Foundation under Contracts Nos. 2021.0174, 2021.0299; Ministry of Development of Turkey under Contract No. DPT2006K-120470; National Research Foundation of Korea under Contract No. NRF-2022R1A2C1092335; National Science and Technology fund of Mongolia; National Science Research and Innovation Fund (NSRF) via the Program Management Unit for Human Resources \& Institutional Development, Research and Innovation of Thailand under Contracts Nos. B16F640076, B50G670107; Polish National Science Centre under Contract No. 2019/35/O/ST2/02907; Swedish Research Council under Contract No. 2019.04595; The Swedish Foundation for International Cooperation in Research and Higher Education under Contract No. CH2018-7756; U. S. Department of Energy under Contract No. DE-FG02-05ER41374.

\bibliographystyle{apsrev4-1}
\bibliography{mybib.bib}

\end{document}